\documentclass[reprint,amsmath,amssymb,aps,prl]{revtex4-1}
\usepackage[utf8]{inputenc}
\usepackage{natbib}
\usepackage{graphicx}
\usepackage{array}
\usepackage{xfrac}
\usepackage{hyperref}
\usepackage{xcolor}

\usepackage{mathtools}

\begin{document}
\title{Asymmetry in repeated isotropic rotations}

\author{Malte Schröder}
\affiliation{Chair for Network Dynamics, Center for Advancing Electronics Dresden (cfaed) and Institute for Theoretical Physics, TU Dresden, 01062 Dresden, Germany}
\affiliation{Cluster of Excellence Physics of Life, TU Dresden, 01062 Dresden, Germany}
\author{Marc Timme}
\affiliation{Chair for Network Dynamics, Center for Advancing Electronics Dresden (cfaed) and Institute for Theoretical Physics, TU Dresden, 01062 Dresden, Germany}
\affiliation{Cluster of Excellence Physics of Life, TU Dresden, 01062 Dresden, Germany}

\begin{abstract}
Random operators constitute fundamental building blocks of models of complex systems yet are far from fully understood. 
Here, we explain an asymmetry emerging upon repeating identical isotropic (uniformly random) operations. 
Specifically, in two dimensions, repeating an isotropic rotation twice maps a given point on the two-dimensional unit sphere (the unit circle) uniformly at random to any point on the unit sphere, reflecting a statistical symmetry as expected. In contrast, in three and higher dimensions, a point is mapped more often closer to the original point than a uniform distribution predicts. Curiously, in the limit of the dimension $d \rightarrow \infty$, a symmetric distribution is approached again. We intuitively explain the emergence of this asymmetry and why it disappears in higher dimensions by disentangling isotropic rotations into a sequence of partial actions.
The asymmetry emerges in two qualitatively different forms and for a wide range of general random operations relevant in complex systems modeling, including repeated continuous and discrete rotations, roto-reflections and general orthogonal transformations.
\end{abstract}

\maketitle

Random operations ubiquitously appear in complex systems models where they often reflect a statistical or approximate symmetry of the real system \cite{wigner67_randomMatrix, may72_will, kottos97_quantum, timme04_speedLimit, jirsa04_will, sarika07_universality, sarika07_random, seba09_parkingDistancePerception, livan18_introduction, moran19_largeEconomyStable}.
Such operations play a special role in physics and are the basic objects of random matrix theory \cite{mehta91_random, livan18_introduction}. Random matrix theory asserts that spectral and statistical properties of complex physical systems are well described by those of random operators given (statistically) the same symmetry. Applications started with Eugene Wigner explaining the spacing statistics of energy levels in atomic nuclei \cite{wigner67_randomMatrix}, and today cover fields as diverse as quantum chaos \cite{kottos97_quantum},  traffic dynamics \cite{seba09_parkingDistancePerception}, economics \cite{moran19_largeEconomyStable} and neurophysics \cite{sompolinsky88_nerualChaos, timme04_speedLimit, wainrib13_complexityRandomNeural} as well as generic complex systems \cite{jirsa04_will, sarika07_universality, sarika07_random}. 
 
Orthogonal transformations, and in particular rotations, are a special class of these operations with fundamental importance across physics, where they reflect rotational invariance resulting from rotational symmetry (isotropy) of the system under consideration. Examples range from  a gravitational potential forcing a planet to revolve around a star and the classical dynamics of the spinning top to the dynamics of isotropic fluids, exactly rotationally invariant spin systems and nearly isotropic superconductors. Furthermore, rotational invariance enters various fundamental theories in physics, for instance the cosmological principle, essentially assuming an isotropic universe, and Noether's theorem, relating rotational invariance to conservation of angular momentum. 

Mathematically, rotations are described by orthogonal matrices $Q\in\mathbb{R}^{d\times d}$ satisfying $Q^{-1}=Q^\textsf{T}$ and $\det{Q}=1$. Each given $Q$ maps one orientation of the unit sphere $S^{d-1}$ in $d$ dimensions to one other orientation, and each point on the sphere to another point on the sphere.
Random \emph{isotropic} rotations appear particularly simple and are characterized by rotation matrices drawn uniformly from the space of all such matrices $Q\in\mathbb{R}^{d\times d}$ defined above. While the theoretical foundations and mathematical descriptions of rotations and related random operations are well established, some paradoxical properties of random rotations still lack an intuitive, descriptive explanation. In this article, we give an intuitive explanation of an asymmetry emerging from the repeated action of isotropic (uniformly random) rotations first presented in 2002 in the context of signal transmission and encoding \cite{marzetta02} and quantitatively analyzed in 2009 using measure theory \cite{eaton09}. By decomposing the isotropic rotation into two sequential elementary rotations, we illustrate the geometric basis for the emergence of this asymmetry for isotropic rotations in particular and orthogonal transformations in general.\\

\begin{figure}
	\centering
	\includegraphics{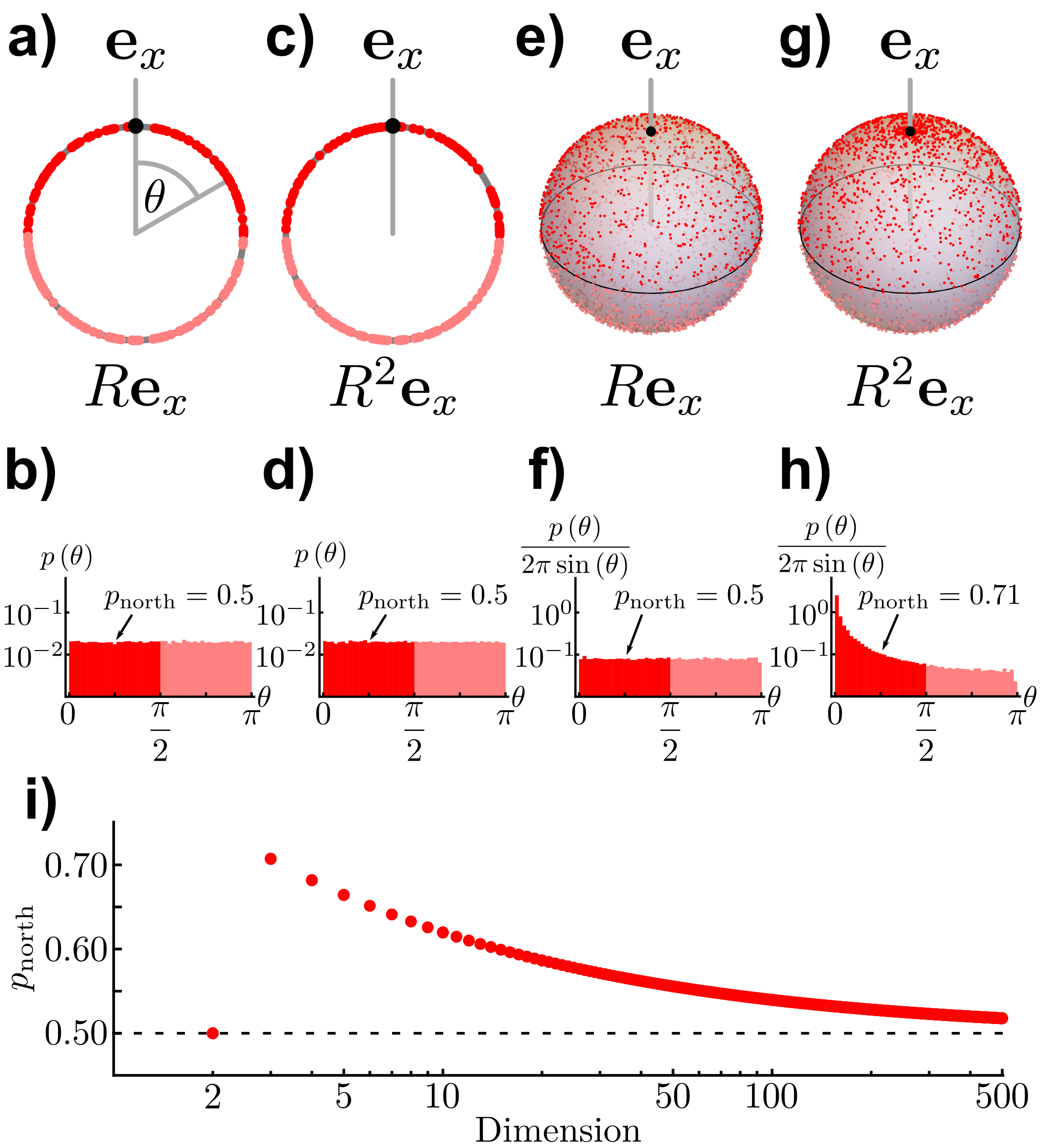}
	\caption{ textbf{Asymmetry from repeated isotropic rotations.}
(a,c,e,g) Realizations of the image of the north pole $\mathbf{e}_x$ under isotropic rotations $R$ and double rotations $R^2$ in two and three dimensions.
(b,d,f,h) Distribution of the angle $\theta$ of the images with respect to $\mathbf{e}_x$. Images on the northern hemisphere are shaded dark red. The distribution is non-uniform for repeated isotropic rotations in three dimensions.
(i) Probability $p_\mathrm{north}$ of finding the image $R^2 \mathbf{e}_x$ on the northern hemispehere (compare \cite{eaton09}). The asymmetry appears first in three dimensions and disappears again as the dimension approaches $d \rightarrow \infty$.
	}
	\label{fig:iso_rot}
\end{figure}

\textit{Asymmetry from repeated isotropic rotations} --- 
Consider a rotation $R$ in $d=2$ dimensions. It maps a point on a circle at angle $\phi$ to an angle $\phi' = \phi + \alpha$. For a rotation $R$ drawn uniformly at random, $\alpha $ is distributed uniformly in $\left[0,2 \pi\right)$. Consequently, after the rotation the point $\phi'=\phi+\alpha \  \mathrm{mod} \ 2\pi$ is distributed uniformly on the circle, 
\begin{equation}
    R\phi \sim \mathrm{Uniform}\left[0, 2\pi\right),
\end{equation}
see Fig.~\ref{fig:iso_rot}(a,b).
Applying the identical rotation $R$ again maps the point to an angle $RR\phi = R^2\phi = \phi + 2 \alpha  \ \mathrm{mod} \ 2\pi$. Also after this second rotation the image 
is uniformly distributed on the circle, \begin{equation}
    R^2\phi  \sim \mathrm{Uniform}\left[0, 2\pi\right),
\end{equation} see Fig.~\ref{fig:iso_rot}(c,d).

\begin{table}
\centering
\begin{tabular}{c|>{\centering\arraybackslash}p{0.65cm}|>{\centering\arraybackslash}p{0.65cm}|>{\centering\arraybackslash}p{0.65cm}|>{\centering\arraybackslash}p{0.65cm}|>{\centering\arraybackslash}p{0.65cm}|>{\centering\arraybackslash}p{0.65cm}|>{\centering\arraybackslash}p{0.65cm}|>{\centering\arraybackslash}p{0.65cm}|>{\centering\arraybackslash}p{0.65cm}|>{\centering\arraybackslash}p{0.65cm}}
	$d$ &  2 & 3 & 4 & 5 & 10 & 20 & 50 & 100 & 500 & $\infty$ \\
	\hline
	$p_\mathrm{north}$ & $\sfrac{1}{2}$ & 0.71 & 0.68 & 0.66 & 0.62 & 0.59 & 0.56 & 0.54 & 0.52 & $\sfrac{1}{2}$ \\
\end{tabular}
\caption{\textbf{Asymmetry from repeated isotropic rotations.} Probability $p_\mathrm{north}(R^2) = \mathrm{Prob}(\mathbf{e}_x^\mathsf{T} R^2  \mathbf{e}_x > 0)$ of finding the image $R^2  \mathbf{e}_x$ of the north pole $\mathbf{e}_x$ on the northern hemisphere after applying the same random isotropic rotation $R$ twice in different dimensions $d$. Decimal values rounded to two significant digits (compare \cite{eaton09}). 
}
\label{tab:p_2}
\end{table}

In dimensions $d \geq 3$, however, the result is qualitatively 
different. Applying a uniformly random (isotropic) rotation $R$ (now denoting a rotation matrix) maps the ``north pole'' $\mathbf{e}_x = \left(1, 0, 0\right)^\mathsf{T} \in \mathbb{R}^3$ to a random point $R \mathbf{e}_x$. By definition, the image is uniformly distributed on the unit sphere [Fig.~\ref{fig:iso_rot}(e,f)].
Applying the same rotation again results in the image $R^2 \mathbf{e}_x$ on the sphere. However, unlike in two dimensions, the image after the second rotation is not uniformly distributed on the sphere. Instead, the image is more likely located on the northern hemisphere, defined as those points $\mathbf{x}$ where $\mathbf{e}_x^\mathsf{T} \mathbf{x} > 0$. This means that the probability $p_\mathrm{north}$ of finding the image on the northern hemisphere is $p_\mathrm{north} = \mathrm{Prob}\left( \mathbf{e}_x^\mathsf{T} R^2  \mathbf{e}_x > 0 \right) > 1/2$ [Fig.~\ref{fig:iso_rot}(g,h)]. Note that the same holds for any point $\mathbf{x}$ and the probability $\mathrm{Prob}\left(\mathbf{x}^\mathsf{T} R^2 \mathbf{x}\right) > 1/2$.

This asymmetry is strongest in three dimensions and decays again as the dimensionality increases [Fig.~\ref{fig:iso_rot}(i) and Tab.~\ref{tab:p_2}]. This phenomenon, termed the \textit{North Pole Problem} \cite{eaton09}, was initially discussed in 2002 in the context of signal transmission and encoding \cite{marzetta02} and mathematically analyzed in 2009 using measure theory \cite{eaton09}. Can we intuitively understand the mechanism behind it? In the following we explicitly construct the repeated isotropic rotation in terms of elementary actions and thereby explain, first, why the asymmetry appears in $d=3$ dimensions, and second, why it decays as $d\rightarrow \infty$.\\

\textit{Disentangling repeated isotropic rotations} --- 
The isotropy of a rotation $R$ is defined by the rotational invariance of its distribution, i.e., the fact that applying any given rotation before or after does not change its distribution, $\mathcal{L}\left(R\right) = \mathcal{L}\left(Q \circ R\right) = \mathcal{L}\left(R \circ Q'\right)$, where $Q$ and $Q'$ are arbitrary rotations and $\mathcal{L}\left(\cdot\right)$ denotes the distribution (``probability density'') of its argument. It follows that the same holds for the distribution of the images $R \mathbf{x}$ when the rotation is applied to any vector $\mathbf{x} \in S^{d-1}$. Since $\mathcal{L}\left(R \mathbf{x}\right) = \mathcal{L}\left(R \circ Q' \mathbf{x}\right)$ and $Q'$ can be any arbitrary rotation, the resulting distribution cannot depend in any way on the original point. Therefore, as intuitively expected, $\mathcal{L}\left(R \mathbf{x}\right)$ must be the distribution reflecting the uniform Lebesgue measure on $S^{d-1}$, as illustrated in Fig.~\ref{fig:iso_rot}(a,b,e,f). As the action of isotropic rotations is independent of the specific initial point $\mathbf{x}$, 
we consider, without loss of generality, the ``north pole'', i.e. the unit vector of the first Cartesian coordinate $\mathbf{e}_x=(1,0,\ldots)^\mathsf{T}\in\mathbb{R}^d$ as our original point.

The first question is now why applying the same random rotation twice results in an asymmetric distribution of images $\mathcal{L}\left(R^2 \mathbf{x}\right)$ in $d \ge 3$ dimensions [Fig.~\ref{fig:iso_rot}(g,h)]? 

To understand the action of an isotropic rotation, we explicitly construct it. Specifically, in three dimensions with Cartesian coordinates $x$,$y$, and $z$, we use the following idea: an isotropic rotation must re-orient the three coordinate axes uniformly to any orientation. Notably, the direction of the $z$-axis is defined by the direction of the other two axes and the right-handedness of the coordinate system. In order to orient the remaining $x$- and $y$-axes, we first fix the new direction of the $x$-axis by uniformly choosing a point $\mathbf{v}$ on the unit sphere and applying the rotation $R_{\mathbf{e}_x \rightarrow \mathbf{v}}\left(\theta_\mathbf{v}\right)$ that maps $\mathbf{e}_x$ to $\mathbf{v}$ in the most direct way, i.e., the rotation with the smallest possible angle $\theta_\mathbf{v} = \arccos\left( \mathbf{e}_x^T \mathbf{v} \right)$ in the plane spanned by $\mathbf{e}_x$ and $\mathbf{v}$.
The $y$-axis can then be oriented by choosing a direction perpendicular to the (new) $x$-axis uniformly at random. Equivalently, we can apply an isotropic rotation in two dimensions around the (new) $x$-axis, i.e., around $\mathbf{v}$. This rotation will map the $y$-axis to a uniformly distributed random direction perpendicular to $\mathbf{v}$. As we saw above, this is simply a rotation around $\mathbf{v}$ (in the subspace $\mathbb{R}^3_{\perp \mathbf{v}}$ orthogonal to $\mathbf{v}$) by an angle $\alpha$ chosen uniformly in $\left[0,2\pi\right)$, which we denote as $R_{\circlearrowleft \mathbf{v}}\left(\alpha\right)$. Together, this defines an isotropic rotation of $S^{2}\subset \mathbb{R}^3$ as
\begin{equation}
R = R_{\circlearrowleft \mathbf{v}} \left(\alpha\right) \circ R_{\mathbf{e}_x \rightarrow \mathbf{v}}\left(\theta_\mathbf{v}\right) \, \label{eq:gamma}
\end{equation}
orienting first the x-axis and then the y-axis to a uniformly chosen orientation of the sphere.
We can visualize the action of the above construction on a globe in 3-dimensional space: 
we first position the globe such that the north pole is pointing in a uniformly random direction and then turn the globe around its north-south axis by a random angle $\alpha$ drawn from a uniform distribution on $[0,2\pi)$. The entire construction selects one orientation of the globe and thus one rotation uniformly from all possible rotations. Importantly, this construction is independent of the explicit choice of $\mathbf{e}_x$ as our north pole.  
Moreover, the same idea can be used recursively in higher dimensions and is, in a more general setting, known from the subgroup algorithm for generating uniform random variables \cite{diaconis87}.\\

\begin{figure}
	\centering
	\includegraphics{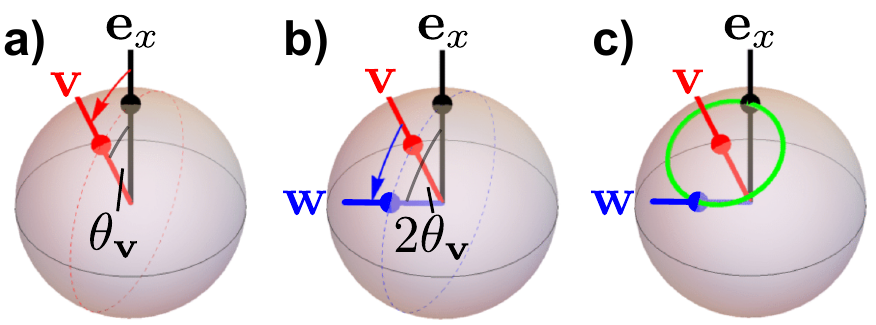}
\caption{
\textbf{Repeated isotropic rotation in three dimensions.} 
Using the explicit construction of an isotropic rotation $R$ Eq.~(\ref{eq:gamma}), we illustrate the action of the repeated rotation $R^2$ on the north pole $\mathbf{e}_x$ step by step [compare (Eq.~\ref{eqn:rot})].
(a) The first application of $R=R_{\circlearrowleft \mathbf{v}} \left(\alpha\right) \circ R_{\mathbf{e}_x \rightarrow \mathbf{v}}\left(\theta_\mathbf{v}\right)$ maps $\mathbf{e}_x$ to a random point $\mathbf{v} = R_{\mathbf{e}_x \rightarrow \mathbf{v}}\left(\theta_\mathbf{v}\right)\mathbf{e}_x$ (red) uniformly distributed on $S^2$, rotating it by an angle $\theta_\mathbf{v}$. The rotation $R_{\circlearrowleft \mathbf{v}} \left(\alpha\right)$ around $\mathbf{v}$ leaves it unchanged.
 (b) The second application first rotates $\mathbf{v}$ again by the same angle $\theta_\mathbf{v}$ to the point $\mathbf{w} = R_{\mathbf{e}_x \rightarrow \mathbf{v}}\left(\theta_\mathbf{v}\right)^2 \mathbf{e}_x$ (blue). 
(c) The final rotation around $\mathbf{v}$ by an angle $\alpha \in \left[0,2\pi\right)$ leaves the angle $\theta_v$ between $\mathbf{w}$ and $\mathbf{v}$ constant, mapping $\mathbf{w}$ to a point uniformly distributed on the circle through $\mathbf{w}$ and $\mathbf{e}_x$ through $\mathbf{v}$ (green). 
For almost all points $\mathbf{v}$ on the sphere, this resulting image $R^2 \mathbf{e}_x$ is more likely on the northern hemisphere. Only if $\mathbf{v}$ is exactly on the equator, defined by $\mathbf{e}_x^T \mathbf{v}=0$, are the images distributed equally between northern and southern hemisphere.
}
	\label{fig:vis}
\end{figure}

Armed with this construction we explain the emergence of asymmetry by explicitly following how each step of 
\begin{equation*}
    R^2 = R_{\circlearrowleft \mathbf{v}} \left(\alpha\right) \circ R_{\mathbf{e}_x \rightarrow \mathbf{v}}\left(\theta_\mathbf{v}\right) \circ R_{\circlearrowleft \mathbf{v}} \left(\alpha\right) \circ R_{\mathbf{e}_x \rightarrow \mathbf{v}}\left(\theta_\mathbf{v}\right)
\end{equation*} 
affects the north pole $\mathbf{e}_x$: (i) The first elementary rotation (the rightmost part) moves the north pole $\mathbf{e}_x$ to its first image $\mathbf{v}$ by the rotation with angle $\theta_\mathbf{v}$ [Fig.~\ref{fig:vis}(a)]. (ii) The second rotation then rotates around $\mathbf{v}$, leaving the image $\mathbf{v}$ of the north pole unchanged. Applying the same rotation again, (iii) $\mathbf{v}$ is first mapped to a new position $\mathbf{w} = R_{\mathbf{e}_x \rightarrow \mathbf{v}}\left(\theta_\mathbf{v}\right)^2 \mathbf{e}_x$, effectively rotating $\mathbf{e}_x$ by an angle $2 \theta_\mathbf{v}$ [Fig.~\ref{fig:vis}(b)]. (iv) The last rotation of $\mathbf{w}$ around $\mathbf{v}$ yields the final second-iterate image $R^2 \mathbf{e}_x$ of the north pole. This final image is a point at angle $\theta_\mathbf{v}$ to $\mathbf{v}$, i.e., a random point uniformly distributed on the circle centered at $\mathbf{v}$ through $\mathbf{w}$. 
We note that both $\mathbf{e}_x$ and $\mathbf{w}$ form the same angle $\theta_\mathbf{v}$ with $\mathbf{v}$, thus both points lie on this circle [Fig.~\ref{fig:vis}(c)]. Overall, we find
\begin{eqnarray}
& &R^2 \mathbf{e}_x  \nonumber \\
&=& \underbrace{R_{\circlearrowleft \mathbf{v}} \left(\alpha\right) \circ R_{\mathbf{e}_x \rightarrow \mathbf{v}}\left(\theta_\mathbf{v}\right)}_{R} \circ \underbrace{R_{\circlearrowleft \mathbf{v}} \left(\alpha\right) \circ R_{\mathbf{e}_x \rightarrow \mathbf{v}}\left(\theta_\mathbf{v}\right)}_{R} \mathbf{e}_x \nonumber\\
            &=& R_{\circlearrowleft \mathbf{v}} \left(\alpha\right) \circ R_{\mathbf{e}_x \rightarrow \mathbf{v}} \left(\theta_\mathbf{v}\right) \mathbf{v} \nonumber\\
			&=& R_{\circlearrowleft \mathbf{v}} \left(\alpha \right) \circ \mathbf{w} \nonumber\\
			&=& R_{\circlearrowleft \mathbf{v}} \left(\alpha'\right) \circ \mathbf{e}_x \,, \label{eqn:rot}
\end{eqnarray}
where $\alpha'=\alpha + \pi \ \textrm{mod}\, 2 \pi$. Since the rotation $R_{\circlearrowleft \mathbf{v}} \left(\alpha\right)$ around $\mathbf{v}$ is isotropic, we can equivalently rotate $\mathbf{e}_x$ by an angle $\alpha'$ drawn from the uniform distribution on $[0,2\pi)$ around the random axis $\mathbf{v}$ and obtain the same distribution, $\mathcal{L}\left(R^2 \mathbf{e}_x\right) = \mathcal{L}\left( R_{\circlearrowleft \mathbf{v}}\left(\alpha'\right) \mathbf{e}_x\right)$. 

The final image is thus uniformly distributed on a circle on the sphere through $\mathbf{e}_x$ centered at $\mathbf{v}$.
Since all these circles cross in the point $\mathbf{e}_x$ independent of $\mathbf{v}$, 
the image of the north pole is more likely to be close to the original direction $\mathbf{e}_x$ on the northern hemisphere than away from it, close to $-\mathbf{e}_x$ on the southern hemisphere. Only if $\mathbf{v}$ is perpendicular to $\mathbf{e}_x$, i.e., if $\mathbf{v}$ lies on the equator, $\mathbf{e}_x$ is rotated along a great-circle of the sphere and the image is located on the northern or southern hemispheres with equal probabilities. This step-by-step construction, Eq.~(\ref{eqn:rot}), explains the emergence of the probabilistic asymmetry in $d=3$ dimensions.\\

\textit{Decay of asymmetry with increasing dimension} --- 
It remains to answer the second open question: why does the asymmetry become less pronounced as the dimension increases? An explicit construction similar to the above explicates isotropic random rotations in arbitrary dimensions $d \geq 3$. As for $d=3$, a point on the sphere $S^{d-1}$ is selected uniformly at random as the first image $\mathbf{v}=R \mathbf{e}_x$ of the north pole $\mathbf{e}_x\in S^{d-1}\subset \mathbb{R}^d$. As the second step, an isotropic rotation of $\mathbf{e}_x$ ``around'' $\mathbf{v}$ (in the subspace $\mathbb{R}^d_{\perp \mathbf{v}}$ transverse to $\mathbf{v}$) is applied. The main difference now is that for $d>3$ the second rotation $R_{\circlearrowleft \mathbf{v}}$ is itself an isotropic rotation of $S^{d-2}$ in $d-1 > 2$ dimensions and thus cannot be parameterized by a single angle $\alpha$. 
Repeating the argument given above, we find an analogous result: we obtain the same distribution when applying $R^2$ as when simply applying an isotropic rotation in $d-1$ dimensions around $\mathbf{v}$. The resulting image of $\mathbf{e}_x$ is distributed on the northern or southern hemispheres with equal probabilities only if $\mathbf{v}$ lies on the equator of the unit sphere $S^{d-1} \subset \mathbb{R}^d$, i.e., if $\mathbf{v}^\mathsf{T} \mathbf{e}_x = 0$. Otherwise, more weight is given to the northern hemisphere.

\begin{figure}
	\centering
	\includegraphics{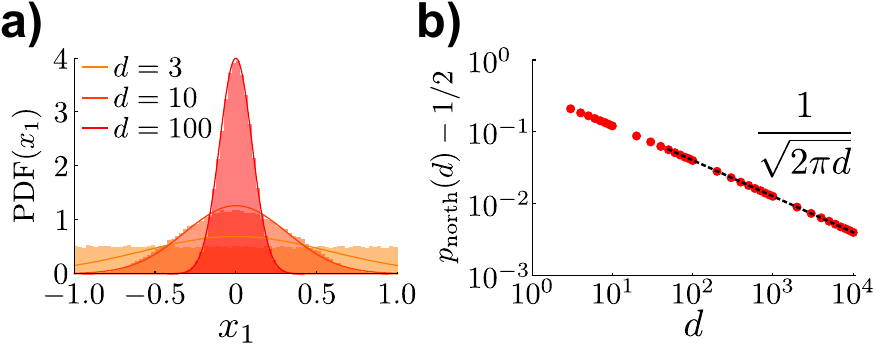}
\caption{
\textbf{$p_\mathrm{north}$ decays with dimension.} 
(a) The probability distribution of the first component $x_1$ of a uniformly random point on a sphere in $d=3,10$ and $100$ dimensions. In higher dimensions, the point is more and more likely to be close to the equator, $x_1 = 0$, and the distribution converges to a normal distribution with mean $0$ and standard deviation $\sqrt{d}$ (solid lines).
(b) Scaling of $p_\mathrm{north}$, the probability that the image $R^2 \mathbf{e}_x$ of the north pole after repeated isotropic rotation is on the northern hemisphere. For large dimensions $d \rightarrow \infty$ the probability $p_\mathrm{north}$ decays to $1/2$ as $p_\mathrm{north} - 1/2 \sim \frac{1}{\sqrt{2 \pi d}}$ [Eq.~(\ref{eq:pnorth})].
}
	\label{fig:scaling}
\end{figure}

Now, similar to the fact that an increasing fraction of the volume of a sphere in $d$ dimensions is located arbitrarily close to its surface as $d$ increases, an increasing fraction of points on the surface is located arbitrarily close to the equator of that sphere as $d$ increases: consider a random vector $\mathbf{x} = \left(x_1, x_2, \dots, x_d\right)^\mathsf{T}$ on the unit sphere $S^{d-1}$ in $d$ dimensions. It has a squared length $x_1^2 + x_2^2 + \dots + x_d^2 = 1$. On average, each $\left<x_i^2\right> = 1/d$ and, for large $d$, the individual coordinates $x_i$ are distributed approximately following a normal distribution with mean $0$ and variance $1/d$ \cite{spruill07_asymptotic, blum19_foundations}, see Fig.~\ref{fig:scaling}(a). In particular, this holds for the first coordinate $x_1$. Consequently, as the dimension increases, the random vector $\mathbf{x}$ is more and more likely to be close to the equator, $x_1^2 \sim 1/d \rightarrow 0$. Thus, in high dimensions, a uniformly chosen direction $\mathbf{v}$ is very likely (almost) perpendicular to $\mathbf{e}_x$ and the final image $R^2 \mathbf{e}_x$ of a repeated isotropic rotation is (almost) equally distributed between northern and southern hemisphere.

The explicit construction explained above suggests a geometric way to exactly calculate the probability $p_\mathrm{north}$ for any dimension $d$ by evaluating the fraction of images of the $d-1$ dimensional isotropic rotation above the equator. 
The detailed integrals of such a geometric construction yield a scaling with dimension of the form
\begin{equation}
    p_\mathrm{north} - 1/2 \sim \frac{1}{\sqrt{2 \pi d}} \label{eq:pnorth}
\end{equation}
as $d \rightarrow \infty$ [Fig.~\ref{fig:scaling}(b)], quantitatively explaining the slow decay observed in Figure \ref{fig:vis}(i). Details of the calculations are presented in the appedix.\\

\begin{figure}[h]
	\centering
	\includegraphics{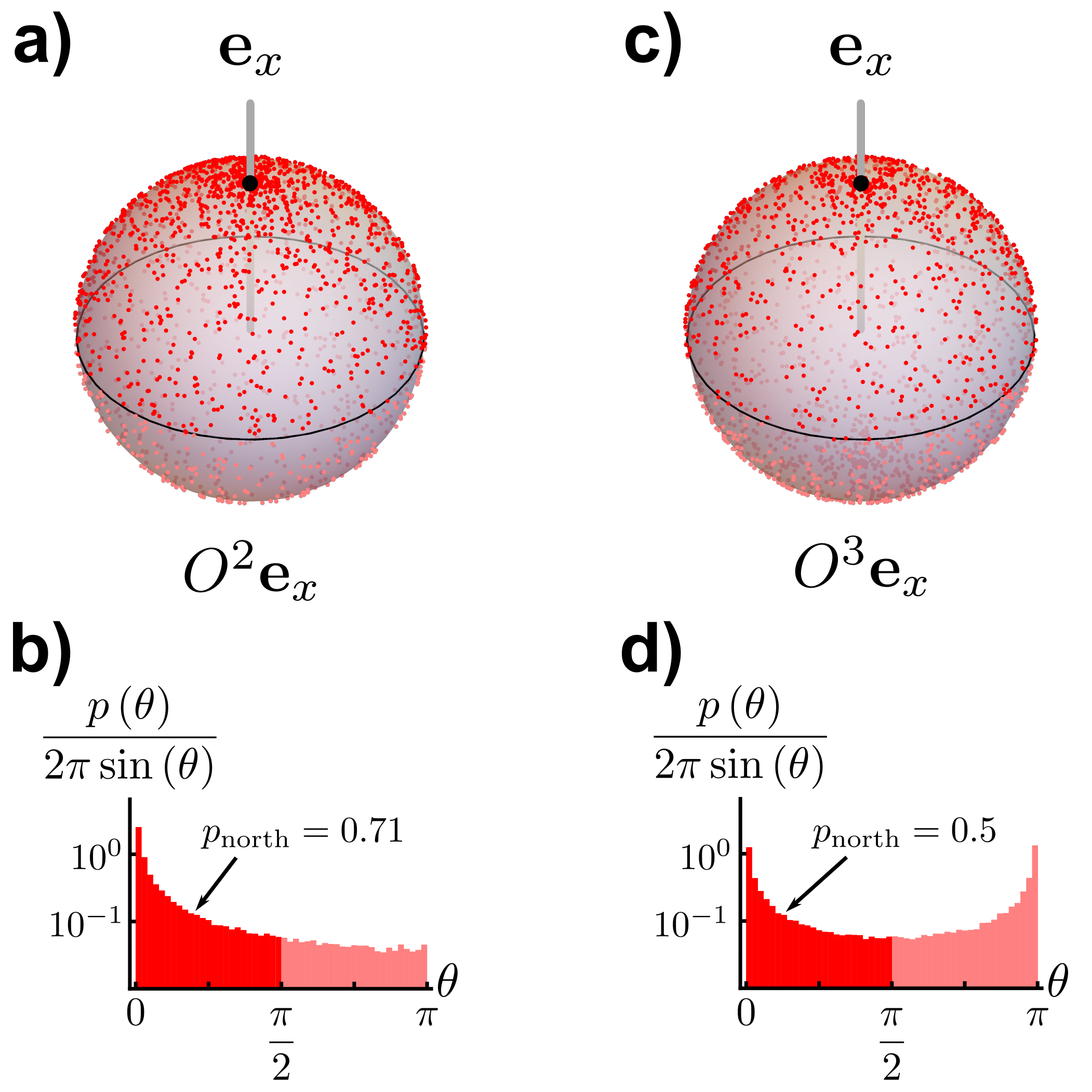}
\caption{\textbf{Repeated random orthogonal transformations.} Repeating other random operations results in different types of asymmetry. (a,b) Repeating the same isotropic orthogonal transformation $O$, i.e., rotation or rotoreflection, twice results in the same distribution of the image as repeating an isotropic rotation $R$ as the reflection inverts itself. (c,d) Repeating $O$ three times results in a 50-50 split between the two hemispheres, but with a non-uniform distribution of the images.
}
\label{fig:orthogonal}
\end{figure}

\textit{Discussion} --- 
In summary, we have explicated and intuitively explained the mechanism underlying an asymmetry in repeated random rotations that emerges in $d=3$ dimensions and disappears again with higher dimensions. Intriguingly, the naively expected 50:50 split of the image of an original point occurs both in dimension $d=2$ and again in the limit $d\rightarrow \infty$, but not in any other dimension.

Such asymmetries emerge not only for continuous rotations. Repeating other random operations, for instance rotations of objects with discrete symmetries such as (hyper)cubic symmetry, will qualitatively face the same breaking of the symmetry. For such discrete random operations, a combinatorial analysis along the arguments detailed above will yield the respective fractions quantifying the asymmetry. Interestingly, the type of breaking of symmetry observed above arises in two qualitatively different versions. First, the 50:50 fraction between hemispheres of images of a single random operation is not preserved by repeating the identical rotation (as for the rotations studied above). Second, even if the image is distributed on the northern and southern hemisphere with a 50:50 split, the distribution of images may not be uniform. This happens, for example, when applying the same random orthogonal transformation $O$ (rotation and/or reflection) three times in $d=3$ dimensions (Fig.~\ref{fig:orthogonal}). 
In general, such (potential) distinctions might be more easily disentangled in systems exhibiting discrete symmetry.
We hope that this article helps to build a better intuition about how basic and ubiquitous symmetry operations act in the physical world around us and about why the three dimensions we live in might be especially interesting.\\

\appendix
\section{Appendix: Geometric calculation of $p_\mathrm{north}$}
The construction explained above suggests a geometric way to exactly calculate the probability $p_\mathrm{north}(d)$ for any dimension $d$ by evaluating the fraction of images of the $d-1$ dimensional isotropic rotation above the equator. These correspond to fractions of the surface of $d-2$ dimensional spheres. For example, in $d=3$ dimensions we need to calculate the fraction of a $d-2 = 1$ dimensional sphere (a circle) that is located above the equator [compare the green circle in Fig.~2(c)]. We then integrate this fraction over all possible directions $\mathbf{v}$ to obtain the final probability. We distinguish two mutually exclusive cases:\\

\begin{figure}
	\centering
	\vspace{3mm}
	\includegraphics{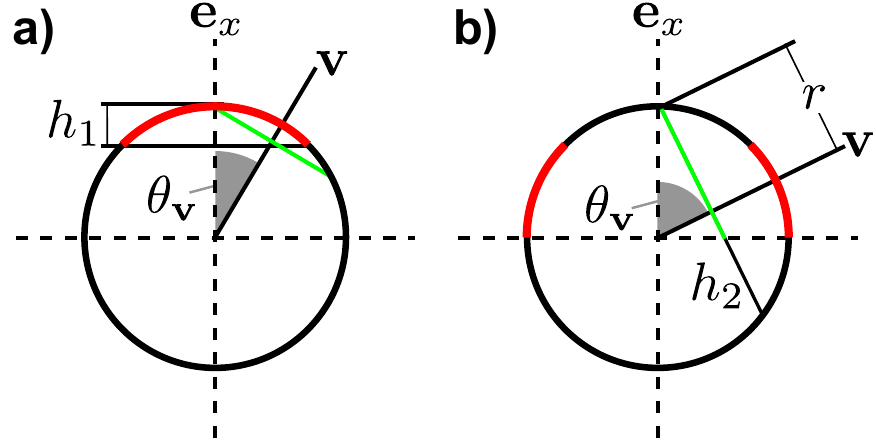}
\caption{
\textbf{Geometric calculation of $p_\mathrm{north}$.} 
Following the argument in the text, we calculate $p_\mathrm{north}$ by integrating over all possible directions $\mathbf{v}$ with the corresponding fraction of images on the northern hemisphere. (a) If $\mathbf{v}$ forms an angle $\theta_\mathbf{v} < \pi/4$ with $\mathbf{e}_x$ (that means if $\mathbf{v}$ is located on the red part of the sphere), the image $R^2 \mathbf{e}_x$ (green) is always on the northern hemisphere [Eq.~(\ref{eq:i_cap}) and (\ref{eq:i_tot})]. (b) Otherwise, if $\pi/4 < \theta_\mathbf{v} < \pi/2$ (red), only a fraction of the final images $R^2\mathbf{e}_x$ lies on the northern hemisphere (green). A fraction corresponding to a $d-2$ dimensional spherical cap with height $h_2$ is mapped to the southern hemisphere [Eq.~(\ref{eq:ii_cap}) and (\ref{eq:ii_tot})]. Both arguments are symmetric with respect to $\mathbf{v}$ on the southern hemisphere.
}
	\label{fig:geo}
\end{figure}

\noindent (i) If $\mathbf{v}$ forms an angle $\theta_\mathbf{v} < \pi/4$ with $\mathbf{e}_x$, all the images of double rotations $R^2$ lie on the northern hemisphere.
This set of vectors $\mathbf{v}$ forms the cap $S^{(d-1)}(\theta_\mathrm{max})$ of the $d-1$ dimensional unit sphere and is defined by its opening angle $\theta_\mathbf{v} < \theta_\mathrm{max} = \pi/4$ or its height $h_1 = 1 - \cos\left(\theta_\mathrm{max}\right)$ [illustrated in Fig.~\ref{fig:geo}(a)]. The cap has an area \cite{li11_hypersperical_caps}
\begin{eqnarray}
    A^{(d-1)}_1 &=& \int_{S^{(d-1)}(\pi/4)} \mathrm{d}A \label{eq:i_cap} \\
    &=& \frac{\pi^{\frac{d}{2}}}{ \Gamma\left( \frac{d}{2} \right)} I_{1/2}\left(\frac{d-1}{2},\frac{1}{2}\right) \,,  \nonumber
\end{eqnarray}
where $\Gamma(x)$ denotes the Gamma function and $I_x(a,b)$ denotes the regularized incomplete Beta function.
Since $\mathbf{v}$ is distributed uniformly on the unit sphere $S^{d-1} \subset \mathbb{R}^d$, we have to weight the area of the cap relative to the total area of the unit sphere
\begin{equation}
    A^{(d-1)}_\mathrm{tot} = \frac{2 \pi^{\frac{d}{2}}}{ \Gamma\left( \frac{d}{2} \right)} \,. \label{eq:i_tot}
\end{equation}
Due to symmetry, the same is true if $\mathbf{v}$ is on the southern hemisphere, $\theta_\mathbf{v}\in(\tfrac{3\pi}{4},\pi]$, giving a factor $2$ in the final evaluation.\\

\noindent (ii) If $\pi/4 \le \theta_\mathbf{v} < \pi/2$, only a fraction of the images will be on the northern hemisphere. This fraction corresponds to the surface of a $d-2$ dimensional sphere with a cap missing. We describe it as one minus the fraction of points on the southern hemisphere, where the points on the southern hemisphere belong to a cap with height $h_2 = 2\sin\left(\theta_\mathbf{v}\right) - \frac{1}{\sin\left(\theta_\mathbf{v}\right)}$ of a sphere with radius $r = \sin\left(\theta_\mathbf{v}\right)$ in $d-2$ dimensions [illustrated in Fig.~\ref{fig:geo}(b)]. This cap has an area \cite{li11_hypersperical_caps}
\begin{equation}
    A^{(d-2)}_2(r,h_2) = \frac{\pi^{\frac{d-1}{2}}}{ \Gamma\left( \frac{d-1}{2} \right)} r^{d-2} I_{\frac{2rh_2-h_2^2}{r^2}}\left(\frac{d-2}{2},\frac{1}{2}\right) \label{eq:ii_cap}
\end{equation}
and we calculate the fraction relative to the total area of the sphere
\begin{equation}
    A^{(d-2)}_\mathrm{tot}(r) = \frac{2 \pi^{\frac{d-1}{2}}}{ \Gamma\left( \frac{d-1}{2} \right)} r^{d-2}\,. \label{eq:ii_tot}
\end{equation}
We again weight each possible $\mathbf{v}$ with respect to the total area of the unit sphere $S^{d-1} \subset \mathbb{R}^d$ [Eq.~(\ref{eq:i_tot})]. As the first case, this argument is also symmetric with respect to the rotation axis $\mathbf{v}$ on the southern hemisphere, $\theta_\mathbf{v}\in[\tfrac{\pi}{2},\tfrac{3\pi}{4})$.

We calculate the total probability by integrating the corresponding fraction over all directions $\mathbf{v}$, where $\mathbf{v}$ is distributed uniformly on the $d-1$ dimensional sphere. The probability of finding the image of $R^2$ on the northern hemisphere is then given by
\begin{widetext}
\begin{eqnarray}
p_\mathrm{north}(d) &=& \underbrace{2\,\frac{A^{(d-1)}_1}{A^{(d-1)}_\mathrm{tot}}}_{\mathrm{case \,\, (i)}}  
    + \underbrace{2\,\frac{1}{A^{(d-1)}_\mathrm{tot}} \int_{\theta_\mathbf{v} \in \left[\frac{\pi}{4}, \frac{\pi}{2}\right)} 1-\frac{A^{(d-2)}_2(r,h_2)}{A^{(d-2)}_\mathrm{tot}(r)} \, \mathrm{d} A}_{\mathrm{case \,\, (ii)}} \nonumber \\
    &=& 2 \, I_{1/2}\left(\frac{d-1}{2},\frac{1}{2}\right) + \frac{2}{A^{(d-1)}_\mathrm{tot}} \int_{\pi/4}^{\pi/2} 1-\frac{1}{2}I_{\frac{2rh_2-h_2^2}{r^2}}\left(\frac{d-2}{2},\frac{1}{2}\right) \mathrm{d}A(\theta_\mathbf{v}) \nonumber\\
    &=& 1 - \frac{2}{A^{(d-1)}_\mathrm{tot}} \int_{\pi/4}^{\pi/2} \frac{1}{2}I_{\frac{2rh_2-h_2^2}{r^2}}\left(\frac{d-2}{2},\frac{1}{2}\right) \mathrm{d}A(\theta_\mathbf{v})\,, \nonumber \\
    &=& 1 - \frac{\Gamma\left(\frac{d}{2}\right)}{\sqrt{\pi} \Gamma\left(\frac{d-1}{2}\right)} \int_{\pi/4}^{\pi/2} I_{\frac{2\sin^2\left(\theta\right) - 1}{\sin^4\left(\theta\right)}}\left(\frac{d-2}{2},\frac{1}{2}\right) \sin^{d-2}\left(\theta\right) \mathrm{d}\theta \approx \frac{1}{2} + \frac{1}{\sqrt{2 \pi d}} \,,\label{eq:pnorth}
\end{eqnarray}
\end{widetext}

where the surface element $\mathrm{d}A(\theta_\mathbf{v}) = \frac{2\pi^{\frac{d-1}{2}}}{ \Gamma\left( \frac{d-1}{2} \right)} \sin\left(\theta_\mathbf{v}\right)^{d-2} \mathrm{d} \theta_\mathbf{v}$ describes the area of the $d-2$ dimensional sphere at angle $\theta_\mathbf{v}$. For example, in $d=3$ dimensions this describes the circumference of a circle of latitude with radius $\sin\left(\theta_\mathbf{v}\right)$ multiplied by $\mathrm{d} \theta_\mathbf{v}$. The last step describes the scaling for large dimensions $d \rightarrow \infty$ obtained via numerical evaluation of the integral. Evaluating this expression gives the results presented in Tab.~I and shown in Fig.~1(i) and Fig.~3(b) (compare also \cite{eaton09}).\\

\begin{acknowledgments}
We gratefully acknowledge support by the Deutsche Forschungsgemeinschaft (DFG, German Research Foundation) under Germany´s Excellence Strategy – EXC-2068 – 390729961 – Cluster of Excellence Physics of Life, the Cluster of Excellence Center for Advancing Electronics at TU Dresden. 
\end{acknowledgments}

\bibliography{randomRotations}

\begin{thebibliography}{19}%
\makeatletter
\providecommand \@ifxundefined [1]{%
 \@ifx{#1\undefined}
}%
\providecommand \@ifnum [1]{%
 \ifnum #1\expandafter \@firstoftwo
 \else \expandafter \@secondoftwo
 \fi
}%
\providecommand \@ifx [1]{%
 \ifx #1\expandafter \@firstoftwo
 \else \expandafter \@secondoftwo
 \fi
}%
\providecommand \natexlab [1]{#1}%
\providecommand \enquote  [1]{``#1''}%
\providecommand \bibnamefont  [1]{#1}%
\providecommand \bibfnamefont [1]{#1}%
\providecommand \citenamefont [1]{#1}%
\providecommand \href@noop [0]{\@secondoftwo}%
\providecommand \href [0]{\begingroup \@sanitize@url \@href}%
\providecommand \@href[1]{\@@startlink{#1}\@@href}%
\providecommand \@@href[1]{\endgroup#1\@@endlink}%
\providecommand \@sanitize@url [0]{\catcode `\\12\catcode `\$12\catcode
  `\&12\catcode `\#12\catcode `\^12\catcode `\_12\catcode `\%12\relax}%
\providecommand \@@startlink[1]{}%
\providecommand \@@endlink[0]{}%
\providecommand \url  [0]{\begingroup\@sanitize@url \@url }%
\providecommand \@url [1]{\endgroup\@href {#1}{\urlprefix }}%
\providecommand \urlprefix  [0]{URL }%
\providecommand \Eprint [0]{\href }%
\providecommand \doibase [0]{http://dx.doi.org/}%
\providecommand \selectlanguage [0]{\@gobble}%
\providecommand \bibinfo  [0]{\@secondoftwo}%
\providecommand \bibfield  [0]{\@secondoftwo}%
\providecommand \translation [1]{[#1]}%
\providecommand \BibitemOpen [0]{}%
\providecommand \bibitemStop [0]{}%
\providecommand \bibitemNoStop [0]{.\EOS\space}%
\providecommand \EOS [0]{\spacefactor3000\relax}%
\providecommand \BibitemShut  [1]{\csname bibitem#1\endcsname}%
\let\auto@bib@innerbib\@empty
\bibitem [{\citenamefont {Wigner}(1967)}]{wigner67_randomMatrix}%
  \BibitemOpen
  \bibfield  {author} {\bibinfo {author} {\bibfnamefont {E.~P.}\ \bibnamefont
  {Wigner}},\ }\href@noop {} {\bibfield  {journal} {\bibinfo  {journal} {SIAM
  Review}\ }\textbf {\bibinfo {volume} {9}},\ \bibinfo {pages} {1} (\bibinfo
  {year} {1967})}\BibitemShut {NoStop}%
\bibitem [{\citenamefont {May}(1972)}]{may72_will}%
  \BibitemOpen
  \bibfield  {author} {\bibinfo {author} {\bibfnamefont {R.~M.}\ \bibnamefont
  {May}},\ }\href@noop {} {\bibfield  {journal} {\bibinfo  {journal} {Nature}\
  }\textbf {\bibinfo {volume} {238}},\ \bibinfo {pages} {413} (\bibinfo {year}
  {1972})}\BibitemShut {NoStop}%
\bibitem [{\citenamefont {Kottos}\ and\ \citenamefont
  {Smilansky}(1997)}]{kottos97_quantum}%
  \BibitemOpen
  \bibfield  {author} {\bibinfo {author} {\bibfnamefont {T.}~\bibnamefont
  {Kottos}}\ and\ \bibinfo {author} {\bibfnamefont {U.}~\bibnamefont
  {Smilansky}},\ }\href@noop {} {\bibfield  {journal} {\bibinfo  {journal}
  {Phys. Rev. Lett.}\ }\textbf {\bibinfo {volume} {79}},\ \bibinfo {pages}
  {4794} (\bibinfo {year} {1997})}\BibitemShut {NoStop}%
\bibitem [{\citenamefont {Timme}\ \emph {et~al.}(2004)\citenamefont {Timme},
  \citenamefont {Wolf},\ and\ \citenamefont {Geisel}}]{timme04_speedLimit}%
  \BibitemOpen
  \bibfield  {author} {\bibinfo {author} {\bibfnamefont {M.}~\bibnamefont
  {Timme}}, \bibinfo {author} {\bibfnamefont {F.}~\bibnamefont {Wolf}}, \ and\
  \bibinfo {author} {\bibfnamefont {T.}~\bibnamefont {Geisel}},\ }\href@noop {}
  {\bibfield  {journal} {\bibinfo  {journal} {Phys. Rev. Lett.}\ }\textbf
  {\bibinfo {volume} {92}},\ \bibinfo {pages} {074101} (\bibinfo {year}
  {2004})}\BibitemShut {NoStop}%
\bibitem [{\citenamefont {Jirsa}\ and\ \citenamefont
  {Ding}(2004)}]{jirsa04_will}%
  \BibitemOpen
  \bibfield  {author} {\bibinfo {author} {\bibfnamefont {V.~K.}\ \bibnamefont
  {Jirsa}}\ and\ \bibinfo {author} {\bibfnamefont {M.}~\bibnamefont {Ding}},\
  }\href@noop {} {\bibfield  {journal} {\bibinfo  {journal} {Phys. Rev. Lett.}\
  }\textbf {\bibinfo {volume} {93}},\ \bibinfo {pages} {070602} (\bibinfo
  {year} {2004})}\BibitemShut {NoStop}%
\bibitem [{\citenamefont {Bandyopadhyay}\ and\ \citenamefont
  {Jalan}(2007)}]{sarika07_universality}%
  \BibitemOpen
  \bibfield  {author} {\bibinfo {author} {\bibfnamefont {J.~N.}\ \bibnamefont
  {Bandyopadhyay}}\ and\ \bibinfo {author} {\bibfnamefont {S.}~\bibnamefont
  {Jalan}},\ }\href@noop {} {\bibfield  {journal} {\bibinfo  {journal} {Phys.
  Rev. E}\ }\textbf {\bibinfo {volume} {76}},\ \bibinfo {pages} {026109}
  (\bibinfo {year} {2007})}\BibitemShut {NoStop}%
\bibitem [{\citenamefont {Jalan}\ and\ \citenamefont
  {Bandyopadhyay}(2007)}]{sarika07_random}%
  \BibitemOpen
  \bibfield  {author} {\bibinfo {author} {\bibfnamefont {S.}~\bibnamefont
  {Jalan}}\ and\ \bibinfo {author} {\bibfnamefont {J.~N.}\ \bibnamefont
  {Bandyopadhyay}},\ }\href@noop {} {\bibfield  {journal} {\bibinfo  {journal}
  {Phys. Rev. E}\ }\textbf {\bibinfo {volume} {76}},\ \bibinfo {pages} {046107}
  (\bibinfo {year} {2007})}\BibitemShut {NoStop}%
\bibitem [{\citenamefont
  {{\v{S}}eba}(2009)}]{seba09_parkingDistancePerception}%
  \BibitemOpen
  \bibfield  {author} {\bibinfo {author} {\bibfnamefont {P.}~\bibnamefont
  {{\v{S}}eba}},\ }\href@noop {} {\bibfield  {journal} {\bibinfo  {journal} {J.
  Stat. Mech.}\ }\textbf {\bibinfo {volume} {2009}},\ \bibinfo {pages} {L10002}
  (\bibinfo {year} {2009})}\BibitemShut {NoStop}%
\bibitem [{\citenamefont {Livan}\ \emph {et~al.}(2018)\citenamefont {Livan},
  \citenamefont {Novaes},\ and\ \citenamefont {Vivo}}]{livan18_introduction}%
  \BibitemOpen
  \bibfield  {author} {\bibinfo {author} {\bibfnamefont {G.}~\bibnamefont
  {Livan}}, \bibinfo {author} {\bibfnamefont {M.}~\bibnamefont {Novaes}}, \
  and\ \bibinfo {author} {\bibfnamefont {P.}~\bibnamefont {Vivo}},\ }\href@noop
  {} {\emph {\bibinfo {title} {Introduction to Random Matrices: Theory and
  Practice}}}\ (\bibinfo  {publisher} {Springer International Publishing},\
  \bibinfo {year} {2018})\BibitemShut {NoStop}%
\bibitem [{\citenamefont {Moran}\ and\ \citenamefont
  {Bouchaud}(2019)}]{moran19_largeEconomyStable}%
  \BibitemOpen
  \bibfield  {author} {\bibinfo {author} {\bibfnamefont {J.}~\bibnamefont
  {Moran}}\ and\ \bibinfo {author} {\bibfnamefont {J.-P.}\ \bibnamefont
  {Bouchaud}},\ }\href@noop {} {\bibfield  {journal} {\bibinfo  {journal}
  {Phys. Rev. E}\ }\textbf {\bibinfo {volume} {100}},\ \bibinfo {pages}
  {032307} (\bibinfo {year} {2019})}\BibitemShut {NoStop}%
\bibitem [{\citenamefont {Mehta}(1991)}]{mehta91_random}%
  \BibitemOpen
  \bibfield  {author} {\bibinfo {author} {\bibfnamefont {M.~L.}\ \bibnamefont
  {Mehta}},\ }\href@noop {} {\emph {\bibinfo {title} {Random Matrices}}}\
  (\bibinfo  {publisher} {Academic Press, New York},\ \bibinfo {year}
  {1991})\BibitemShut {NoStop}%
\bibitem [{\citenamefont {Sompolinsky}\ \emph {et~al.}(1988)\citenamefont
  {Sompolinsky}, \citenamefont {Crisanti},\ and\ \citenamefont
  {Sommers}}]{sompolinsky88_nerualChaos}%
  \BibitemOpen
  \bibfield  {author} {\bibinfo {author} {\bibfnamefont {H.}~\bibnamefont
  {Sompolinsky}}, \bibinfo {author} {\bibfnamefont {A.}~\bibnamefont
  {Crisanti}}, \ and\ \bibinfo {author} {\bibfnamefont {H.~J.}\ \bibnamefont
  {Sommers}},\ }\href@noop {} {\bibfield  {journal} {\bibinfo  {journal} {Phys.
  Rev. Lett.}\ }\textbf {\bibinfo {volume} {61}},\ \bibinfo {pages} {259}
  (\bibinfo {year} {1988})}\BibitemShut {NoStop}%
\bibitem [{\citenamefont {Wainrib}\ and\ \citenamefont
  {Touboul}(2013)}]{wainrib13_complexityRandomNeural}%
  \BibitemOpen
  \bibfield  {author} {\bibinfo {author} {\bibfnamefont {G.}~\bibnamefont
  {Wainrib}}\ and\ \bibinfo {author} {\bibfnamefont {J.}~\bibnamefont
  {Touboul}},\ }\href@noop {} {\bibfield  {journal} {\bibinfo  {journal} {Phys.
  Rev. Lett.}\ }\textbf {\bibinfo {volume} {110}},\ \bibinfo {pages} {118101}
  (\bibinfo {year} {2013})}\BibitemShut {NoStop}%
\bibitem [{\citenamefont {Marzetta}\ \emph {et~al.}(2002)\citenamefont
  {Marzetta}, \citenamefont {Hassibi},\ and\ \citenamefont
  {Hochwald}}]{marzetta02}%
  \BibitemOpen
  \bibfield  {author} {\bibinfo {author} {\bibfnamefont {T.~L.}\ \bibnamefont
  {Marzetta}}, \bibinfo {author} {\bibfnamefont {B.}~\bibnamefont {Hassibi}}, \
  and\ \bibinfo {author} {\bibfnamefont {B.~M.}\ \bibnamefont {Hochwald}},\
  }\href@noop {} {\bibfield  {journal} {\bibinfo  {journal} {IEEE Transactions
  on Information Theory}\ }\textbf {\bibinfo {volume} {48}},\ \bibinfo {pages}
  {942} (\bibinfo {year} {2002})}\BibitemShut {NoStop}%
\bibitem [{\citenamefont {Eaton}\ and\ \citenamefont
  {Muirhead}(2009)}]{eaton09}%
  \BibitemOpen
  \bibfield  {author} {\bibinfo {author} {\bibfnamefont {M.~L.}\ \bibnamefont
  {Eaton}}\ and\ \bibinfo {author} {\bibfnamefont {R.~J.}\ \bibnamefont
  {Muirhead}},\ }\href@noop {} {\bibfield  {journal} {\bibinfo  {journal}
  {Stat. Probab. Lett.}\ }\textbf {\bibinfo {volume} {79}},\ \bibinfo {pages}
  {1878} (\bibinfo {year} {2009})}\BibitemShut {NoStop}%
\bibitem [{\citenamefont {Diaconis}\ and\ \citenamefont
  {Shahshahani}(1987)}]{diaconis87}%
  \BibitemOpen
  \bibfield  {author} {\bibinfo {author} {\bibfnamefont {P.}~\bibnamefont
  {Diaconis}}\ and\ \bibinfo {author} {\bibfnamefont {M.}~\bibnamefont
  {Shahshahani}},\ }\href@noop {} {\bibfield  {journal} {\bibinfo  {journal}
  {Probab. Eng. Inf. Sci.}\ }\textbf {\bibinfo {volume} {1}},\ \bibinfo {pages}
  {15} (\bibinfo {year} {1987})}\BibitemShut {NoStop}%
\bibitem [{\citenamefont {Spruill}(2007)}]{spruill07_asymptotic}%
  \BibitemOpen
  \bibfield  {author} {\bibinfo {author} {\bibfnamefont {M.}~\bibnamefont
  {Spruill}},\ }\href@noop {} {\bibfield  {journal} {\bibinfo  {journal} {Elec.
  Comm. in Probab.}\ }\textbf {\bibinfo {volume} {12}},\ \bibinfo {pages} {234}
  (\bibinfo {year} {2007})}\BibitemShut {NoStop}%
\bibitem [{\citenamefont {Blum}\ \emph {et~al.}(2018)\citenamefont {Blum},
  \citenamefont {Hopcroft},\ and\ \citenamefont {Kannan}}]{blum19_foundations}%
  \BibitemOpen
  \bibfield  {author} {\bibinfo {author} {\bibfnamefont {A.}~\bibnamefont
  {Blum}}, \bibinfo {author} {\bibfnamefont {J.}~\bibnamefont {Hopcroft}}, \
  and\ \bibinfo {author} {\bibfnamefont {R.}~\bibnamefont {Kannan}},\
  }\href@noop {} {\emph {\bibinfo {title} {Foundations of data science}}}\
  (\bibinfo  {publisher} {Preprint of a textbook},\ \bibinfo {year} {2018})\
  \bibinfo {note} {available at
  https://www.cs.cornell.edu/jeh/book.pdf}\BibitemShut {NoStop}%
\bibitem [{\citenamefont {Li}(2011)}]{li11_hypersperical_caps}%
  \BibitemOpen
  \bibfield  {author} {\bibinfo {author} {\bibfnamefont {S.}~\bibnamefont
  {Li}},\ }\href@noop {} {\bibfield  {journal} {\bibinfo  {journal} {Asian J.
  Math. Stat.}\ }\textbf {\bibinfo {volume} {4}},\ \bibinfo {pages} {66}
  (\bibinfo {year} {2011})}\BibitemShut {NoStop}%
\end{thebibliography}%

\end{document}